# Ya. Kalinovskiy, Yu. Boiarinova
## Method for representing an exponent in a fifth-dimensional hypercomplex number systems using a hypercomplex computing software

For the first time, a constructive definition of nonlinear transcendental functions from hypercomplex variables (HV) was proposed by V.R. Hamilton, one of the creators of hypercomplex number systems(HNS). In [1], he proposes to define the exponential function of a quaternionic variable as the sum of a power series:

$$\text{Exp}(q) = \sum_{s=0}^{\infty} \frac{q^s}{s!}. \quad (1)$$

Over time, this approach was generalized to other transcendental functions of a HV: trigonometric sine and cosine, hyperboic sine and cosine, and others, and is now generally accepted [2]. In many works, studies were carried out on the construction of an image of the exponential function of the quaternion. For this, various methods are used, which are based on the symmetry properties of quaternions [2,3]. Images of other transcendental functions of the quaternion are also constructed: logarithmic function, trigonometric sine and cosine [3].

These functions have found use not only in scientific applications (physics, mechanics), but also in technical ones(orientation of a rigid body in space, gyroscopy, robotics and others).

The problem of constructing images of nonlinear functions of a HV is reduced to their definition from the point of view of the structure of computations over the hypercomplex argument and their image in the form of a hypercomplex function:

$$F(X) = \sum_{i=1}^{n} f_i(x_1, \ldots, x_n) e_i, \quad (2)$$

where $X = \sum_{i=1}^{n} x_i e_i$, $e_i$, - basic elements of a hypercomplex number system Γ, $f_i(x_1, \ldots, x_n)$, $i = 1, \ldots, n$ - real functions.

Knowledge of the laws of performing algebraic operations in HNS allows, when defining linear or nonlinear functions, to construct an image of them in the form of a hypercomplex function (2). The construction of an image of transcendental functions of a hypercomplex argument is reduced to the transformation of series of type (1) into (2). In some simple cases, this can be done directly. But in the general case, it requires the development of special methods [4].

**Construction of an exponential representation of a HV using a system of differential equations associated with the HNS.** It is very difficult to construct a representation of the exponent of a HV directly from (1) and (2), especially when constructing exponents in noncanonical HNS and HNS of high dimensions.

The proposed generalized construction method [4], which can be applied to any finite-dimensional HNS, is as follows.

Let us denote the column vectors composed of the components of hypercomplex variables by capital Latin letters with a bar, for example, $\overline{X} = (x_1, \ldots, x_n)^T$. The exponent representation in the system $\Gamma(e, n)$ in a variable $M \in \Gamma(e, n)$, which we will denote $\text{Exp}(M)$ is a particular solution of the hypercomplex linear differential equation

$$\dot{X} = MX \quad (3)$$

with the initial condition

$$\text{Exp}(0) = \varepsilon, \quad (4)$$

where ε is a unit element of the system.

To construct a solution of equation (3), it must be represented in vector-matrix form:

$$\dot{\overline{X}} = (\dot{x}_1, \ldots, \dot{x}_n)^T,$$

and the column vector $\overline{MX}$ obtained from the hypercomplex number $MX$ can be represented as the matrix product of some matrix $\Psi$ of size $n \times n$, the elements of which are linear combinations of the components of the hypercomplex number $M$, by the column vector $\bar{X}$:
$$\overline{MX} = \Psi \bar{X}.$$
Then the hypercomplex equation (3) is transformed into a system of n linear differential equations of the first order in real variables, which is called the associated (with a given HNS) system of linear differential equations:
$$\dot{\bar{X}} = \Psi \bar{X}. \qquad (5)$$
Thus, the characteristic numbers $\lambda_1, \ldots, \lambda_n$ will depend on the hypercomplex number $M$.

General solutions $\bar{X}(t, C_1, \ldots, C_n, \overline{M})$ of system (5) depend on n arbitrary constants and components of the hypercomplex number $M$. The value of arbitrary constants are set using the initial condition (4). The components of the column vector $\bar{X}$ of the solution and will be the components of the exponent of the hypercomplex number $M$:
$$\mathrm{Exp}(M) = \sum_{i=1}^{n} \bar{x}_i\, e_i.$$

The method for constructing representations of the exponential in a HV using an associated system of linear differential equations can be easily formalized for constructing algorithms and programs in symbolic computation systems. In this process, the main obstacle is the cumbersomeness of hypercomplex expressions in symbolic form, especially with high dimensions of the HNS. To avoid these difficulties, it is required to develop and use special tools for performing calculations with hypercomplex expressions.

**Main information of hypercomplex computing software.** Almost all computer algebra systems (CAS) have facilities for operating with complex numbers in symbolic or numerical forms. Some SKA, namely: Matlab, Mathcad, Mathematica, Maple, MuPAD, S-PLUS, etc. [5, 6], have facilities for operating with a limited set of HNS (quaternions, octonions, Clifford algebras and some others). The list of operations in these CAS is small. This situation requires the creation of special algorithmic software for solving specific problems using HNS. Therefore, in recent years, in the department of special modeling tools, work has been carried out on the creation of algorithmic and software tools for modeling problems using hypercomplex numerical systems. These tools are a package of programs and procedures created on the basis of CAS Maple. The most complete description of this package is presented in [7].

Since the Maple computer algebra system allows you to create specialized packages of various computational procedures, the created hypercomplex computing software (HCS) is a package with an identifier, HCS can be called, attached to the program and transported to other computers. From the HCS procedures, it is possible to form computation programs using the tools of the Maple algorithmic language.

Much attention was paid to the methods and structures of data presentation in the development of the CCP. As noted above, the PCGO is designed to operate with data in a hypercomplex form:
$$A = a_1 e_1 + a_2 e_2 + \cdots + a_n e_n \qquad (7)$$
where n - dimension of HNS, $a_i$ - algebraic expressions, $e_i$ - elements of the basis of the HNS ("imaginary units"). This form of the hypercomplex number will be called natural. It is rather inconvenient to operate with hypercomplex numbers in natural form. This is due to the fact that various operations are performed with the coefficients of the basic elements that need to be isolated and identified

Consider such a simple operation as the addition of hypercomplex numbers. Let B another number of the same form as (7). If used the addition operation in Maple, we get the following:
$$C = A + B = a_1 e_1 + a_2 e_2 + \cdots + a_n e_n + b_1 e_1 + b_2 e_2 + \cdots + b_n e_n$$

The expression on the right-hand side of (7) is not a natural form of a hypercomplex number. In it, you need to make the reduction of similar terms. The Maple system has a command for converting similar terms (*collect*), but when using it, you must specify the variable by which the similar ones are converted. Therefore, in this case, you will have to use the *collect* command n times, specifying each time with which variable the similar is being cast:
$$C = (collect(\ldots(collect(collect(\ldots(A+B), e_1), \ldots), e_n)) =$$
$$= (a_1 + b_1)e_1 + (a_2 + b_2)e_2 + \cdots + (a_n + b_n)e_n \qquad (8)$$
As you can see, (8) is a complex design, especially at large dimensions of the HNS. There are a lot of such inconveniences associated with using the natural form of representing hypercomplex numbers.

At the same time, the Maple system has tools that allow you to get rid of these and many other inconveniences associated with using the natural form of representing hypercomplex numbers. The fact is that in the natural form of representing hypercomplex numbers, only the coefficients of the basis elements and their ordinal number in the image of the hypercomplex number are important, that is, the hypercomplex number can be represented as a vector. However, the vector-matrix form is not suitable here due to the fact that the components of the matrix and the vector must be of the same type. At the same time, in the Maple system, such a form of data presentation as a list is a list - an ordered set of different types of data. To operate with data in the format of lists in Maple, there are numerous commands that allow you to specify a list(*list* an ordered set of different types of data). To operate with list data in Maple, there are numerous commands that allow you to specify a list, determine its length, make two lists of the same length, determine a list member by its serial number, multiply all list members by any expression, and more.

Representation of hypercomplex numbers in the form of a list is called the list or internal representation of hypercomplex numbers. Thus, we will use the representation $A = [a_1, a_2, \ldots a_n]$. Then the sum of two numbers will be determined:
$$C = A + B = [a_1, \ldots a_n] + [b_1, \ldots, b_n] = [a_1 + b_1, \ldots, a_n + b_n]$$
that is, the conversion of such symbolic coefficients in accordance with their ordinal numbers in numbers is performed by the automatic internal means of Maple.

Thus, the representation of hypercomplex numbers in a list format greatly simplifies software development. However, such a solution requires the presence in the HCS of procedures for the reciprocal transformation of the natural and internal forms of representation of hypercomplex numbers. Moreover, it is advisable to perform some actions on numbers in natural form. In this regard, in many HCS procedures, it is assumed that the result of calculations is output in the form of a list of two elements: the first element is the result in list form, the second is in natural form. So, for example, this is how the procedure for generating a hypercomplex number of the eighth dimension looks like:

> $A = \text{HNSnumber}(8, a, e)$:
> $A[1]$
$$[a_1, a_2, a_3, a_4, a_5, a_6, a_7, a_8]$$
> $A[2]$
$$a_1 e_1 + a_2 e_2 + a_3 e_3 + a_4 e_4 + a_5 e_5 + a_6 e_6 + a_7 e_7 + a_8 e_8$$

It also turned out to be advisable to give the list format and more complex hypercomplex structures. Thus, the Kelley table of multiplication of basic elements is represented by a three-level list structure: the upper level consists of a list of table rows, the second nested level is a list of table elements, the third, lowest level is a list of structural constants of one cell of the Kelly table.

**Characteristics of the class of investigated HNS.** In [12] it is indicated: "The problem of studying arbitrary algebraic structures is very general, so that it is of real value. Therefore, it is considered under various natural restrictions". Since the class of different HNS is very wide, in this paper we will restrict ourselves to considering such HNS, which are group algebras [8], the basis elements of which form groups G isomorphic to additive groups $Z_n$ of classes of residues modulo integer $n$:
$$G \simeq Z_n = Z/nZ.$$

Group operation, and the law of composition of basic elements, is expressed by the formula
$$e_i \cdot e_j = e_{(i+j)(\bmod\ n)}, i,j = 0,1,\ldots,n-1 \quad (9)$$
where n is the dimension of the investigated HNS. As follows from (9), such HNS are commutative and associative. This is what determines the practical value of their use for modeling.

Since it is more convenient to determine the indices of basic elements from 1 to n, then (9) is transformed into the formula
$$e_i \cdot e_j = e_{(i+j-2)(\bmod\ n)+1}, \, i,j = 1,\ldots,n.$$
It is necessary to notice that at such $n \le 4$ $n = 2,3$ HNS are already investigated.

Consider the case n = 4. We obtain a HNS $G_{47}$ according to the HCS classification. The Keli table of which has the form:

| $G_{47}$ | $e_1$ | $e_2$ | $e_3$ | $e_4$ |
|---|---|---|---|---|
| $e_1$ | $e_1$ | $e_2$ | $e_3$ | $e_4$ |
| $e_2$ | $e_2$ | $e_3$ | $e_4$ | $e_1$ |
| $e_3$ | $e_3$ | $e_4$ | $e_1$ | $e_2$ |
| $e_4$ | $e_4$ | $e_1$ | $e_2$ | $e_3$ |

This case is not described in the literature, but the exponent representation is quite easy to construct "by hand".

It can be shown that the multiplicative semigroup of the HNS $G_{47}$ is isomorphic to the subgroup $V = \{e, (1234), (13)(24), (1432)\}$, which is a normal subgroup of the symmetric group $S_4$ : $V \triangleleft S_4$.

Characteristic equation (6) has two real and a pair of complex conjugate roots:
$$\lambda_{1,2} = m_1 + m_3 \pm (m_2 + m_4), \lambda_{3,4} = m_1 - m_3 \pm i(m_2 - m_4).$$
From this fact, we can conclude that the HNS has at least two isomorphic systems [9]: $R \oplus R \oplus C$ and $W \oplus C$. Their Kelly tables look like this:

| $R \oplus R \oplus C$ | $e_1$ | $e_2$ | $e_3$ | $e_4$ |
|---|---|---|---|---|
| $e_1$ | $e_1$ | 0 | 0 | 0 |
| $e_2$ | 0 | $e_3$ | 0 | 0 |
| $e_3$ | 0 | 0 | $e_3$ | $e_4$ |
| $e_4$ | 0 | 0 | $e_4$ | $-e_3$ |

| $W \oplus C$ | $e_1$ | $e_2$ | $e_3$ | $e_4$ |
|---|---|---|---|---|
| $e_1$ | $e_1$ | $e_2$ | 0 | 0 |
| $e_2$ | $e_2$ | $e_1$ | 0 | 0 |
| $e_3$ | 0 | 0 | $e_3$ | $e_4$ |
| $e_4$ | 0 | 0 | $e_4$ | $-e_3$ |

If we introduce the notation $\alpha_{1,2} = m_1 \pm m_3$, $\beta_{1,2} = m_3 \pm m_4$, then the exponent representation in the system $G_{47}$ looks like this:
$$\mathrm{Exp}(m_1 e_1 + m_2 e_2 + m_3 e_3 + m_4 e_4) =$$
$$= \frac{1}{2}[(e^{\alpha_1}\cosh\beta_1 + e^{\alpha_2}\cos\beta_2)e_1 + (e^{\alpha_1}\sinh\beta_1 + e^{\alpha_2}\sin\beta_2)e_2 +$$
$$+ (e^{\alpha_1}\cosh\beta_1 - e^{\alpha_2}\cos\beta_2)e_3 + (e^{\alpha_1}\sinh\beta_1 - e^{\alpha_2}\sin\beta_2)e_4]$$
HNS $G_{47}$ is commutative and associative.

In connection with the above, the goal of the work can be formulated as follows: the solution of some algorithmic problems arising in the modeling of representations of the exponential function in the fifth-dimensional HNS.

**Construction of the exponent in the HNS $G_{51}$ dimension** n = 5. Actions with hypercomplex expressions, especially in high-dimensional HNS, and even in symbolic form, are very cumbersome. Therefore, we will carry out all calculations and transformations in the fifth-dimensional state emergency system using the package of hypercomplex calculations of the HCS. This will both reduce the time spent and avoid mistakes. Our reasoning will be accompanied by fragments of programs in the CAS Maple language environment using the means of HCS.

First of all, it is necessary to call from the storage of HNS LibHNS () the $G_{51}$ HNS and display its Keli table:

$G51 := \text{SearchHNS}(G51, \text{LibHNS}(\ ))$:
$\text{VizHNS}(G51, e)$

$$\begin{vmatrix} e_1 & e_2 & e_3 & e_4 & e_5 \\ e_2 & e_3 & e_4 & e_5 & e_1 \\ e_3 & e_4 & e_5 & e_1 & e_2 \\ e_4 & e_5 & e_1 & e_2 & e_3 \\ e_5 & e_1 & e_2 & e_3 & e_4 \end{vmatrix}$$

To construct the matrix of the right-hand sides (3), it is necessary to perform hypercomplex multiplications and transform the hypercomplex equations (3) into a system of real equations, after which one can directly construct the matrix and find its characteristic numbers. By means HCS of it looks like this:

$M := \text{HNSnumber}(5, m, e): X := \text{HNSnumber}(5, x, e)$:
$MX := \text{inMulti}(X[1], M[1], G51)$:
$\text{Pr: Matrix}(5)$:
**for** i **from** 1 to 5 **do** $s2[i] := \text{convert}(MX[i], \text{list})$ **od**:
**for** k **from** 1 to 5 **do for** i **from** 1 to 5 **do** $j := \text{op}(\text{convert}(s2[k][i], \text{list}[1]))$:
$\Pr[k, j] := \text{convert}(s2[k][i], \text{list})[2]$: **od: od**
Pr,

$$\begin{vmatrix} m_1 & m_5 & m_4 & m_3 & m_2 \\ m_2 & m_1 & m_5 & m_4 & m_3 \\ m_3 & m_2 & m_1 & m_5 & m_4 \\ m_4 & m_3 & m_2 & m_1 & m_5 \\ m_5 & m_4 & m_3 & m_2 & m_1 \end{vmatrix}$$

$\lambda := \text{Eigenvalues}(\Pr)$:

Let us call the set of characteristic numbers for a particular HNS its spectrum. The first root of the spectrum is clearly real: $\lambda_1 = m_1 + m_2 + m_3 + m_4$ and the other four roots form two pairs of the following type:

$$a \pm \sqrt{b}. \tag{10}$$

For the second and third roots

$$a = m_1 - \frac{1}{4}m_5 - \frac{1}{4}m_2 - \frac{1}{4}m_3 - \frac{1}{4}m_4,$$

$$b = (-2\sqrt{5}m_2^2 - 10m_2^2 + 4\sqrt{5}m_5m_2 + 20m_5m_2 - 8\sqrt{5}m_3m_2 + 8\sqrt{5}m_4m_2 - 10m_5^2 \\ + 2\sqrt{5}m_4^2 + 2\sqrt{5}m_3^2 - 8\sqrt{5}m_4m_5 - 10m_4^2 - 2\sqrt{5}m_5^2 - m_3^2 + 20m_3m_4 \\ + 8\sqrt{5}m_3m_5 - 4\sqrt{5}m_3m_4)/4,$$

If a - the number is real, then the sign of b depends on whether the roots (10) are real or complex. It is not known whether b will have a permanent sign. If $b \geq 0$ then this root is a real number, like the sum and difference of two real numbers. If $b < 0$ then the root is complex.

However, in appearance it is very difficult to judge the sign of the expression. It can have different signs depending on the values of the coefficients. Note, however, that if this is the case, then this will indicate an error in the calculations, because, as shown in [9], the type of the spectrum does not change with a linear transformation of the HNS basis.

The expression b is a quadratic form in 4 variables $-m_2, m_3, m_4, m_5$, the matrix of which in this case has the form:

$$KI = \begin{vmatrix} -2\sqrt{5}-10 & -4\sqrt{5} & 4\sqrt{5} & 2\sqrt{5}+10 \\ -4\sqrt{5} & 2\sqrt{5}-10 & -2\sqrt{5}+10 & 4\sqrt{5} \\ 4\sqrt{5} & -2\sqrt{5}+10 & 2\sqrt{5}-10 & -4\sqrt{5} \\ 2\sqrt{5}+10 & 4\sqrt{5} & -4\sqrt{5} & -2\sqrt{5}-10 \end{vmatrix}$$

Quadratic forms can be sign-definite, that is, their values take a certain sign (+ or -) for any values of the arguments of the quadratic form, or sign-alternating otherwise. There are various criteria for determining the type of a quadratic form [11]. The main criterion is Sylvester's criterion, according to which the signs of the major minor of a matrix of a quadratic form are studied. In this case, all major minors are zero. Therefore, it is necessary to use the criterion of signs of the eigenvalues of a matrix of quadratic form.

Using the package $LinearAlgebra$ we determine the eigenvalues of a matrix of quadratic form with procedure $Eigenvalues$: $|0,0,0,-40|$. All four eigenvalues are non-positive, that is, the quadratic form is non-positive and it may be zero. Thus, these two characteristic numbers of a matrix are a complex conjugate pair.

This analysis is rather complicated. And by means of HCS it is very simple, because there is a procedure *ARoot*, which determines the sign of an expression by factoring it. Then we will receive

$$b = -\frac{2\sqrt{5}+10}{64}(-2m_5 + 2m_2 - \sqrt{5}m_3 - \sqrt{5}m_4 - m_3 + m_4)^2 < 0,$$

whence it follows that the characteristic numbers of a matrix are a complex conjugate pair.

Similarly, we can define that the characteristic numbers $\lambda_4, \lambda_5$ are also a complex conjugate pair. The last four characteristic numbers

$$\lambda_{2,3,4,5} = (\frac{1}{4}(m_2 + m_3 + m_4 + m_5) \pm \frac{\sqrt{5}}{4}(m_2 - m_3 - m_4 + m_5)) \pm I(\pm 2m_2 \mp 2m_5 \mp m_3$$
$$\pm m_4 + \sqrt{5}(m_3 - m_4))\frac{\sqrt{10 \pm \sqrt{5}}}{8}$$

Thus, the matrix of the right-hand side of the associated system of linear differential equations for the HNS $G_{51}$ has one real characteristic number and two different pairs of complex conjugates. Therefore, it can be determined that the HNS $G_{51}$ is isomorphic to the direct sum of two systems of complex numbers C and a system of real numbers R:

$$G_{51} \cong R \oplus C \oplus C$$

The multiplicative group HNS $G_{51}$, which we denote by $G_{51*}$, is isomorphic to such a subgroup of the symmetric group $S_5$:

$$V = \{e, (12345), (13524), (14253), (15432)\} \subset S_5$$

Since the subgroup V consists only of odd cycles, that is, of even permutations, it is a subgroup of an alternating subgroup $A_5 \subset S_5$.

For further construction of the exponent, it is necessary to construct particular solutions of the system of equations (5) for all characteristic numbers. Private solutions for $\lambda_1$ should look like:

$$Xp_{1i} = C_{1i}e^{\lambda_1 t}, i = 1,\dots,5. \qquad (11)$$

After substitution in (5) and simplification, a linear system will be obtained, from which the values of arbitrary constants $C_{1i} = idem, i = 1,\dots,5$ can be obtained. That is only one arbitrary constant will remain.

For the first pair of complex-conjugate characteristic numbers $\lambda_{2,3}$, particular solutions should be of the form:

$$Xp_{2i} = e^{re\lambda_2 t}(C_{2i}\cos(im\lambda_2 \cdot t) + F_{2i}\sin(im\lambda_2 \cdot t)), i = 1,\dots,5. \qquad (12)$$

Substitution of these solutions into equation (3) using the HCS *Const1* procedure gives a system of 5 real linear equations. Since these equations are very cumbersome, we present only one of them for illustration:

$$-\frac{1}{4}m_2F_{2,1} - m_5F_{2,2} - \frac{1}{4}m_4F_{2,1} - m_4F_{2,3} - \frac{1}{4}m_3F_{2,1} - m_3F_{2,4} - \frac{1}{4}m_5F_{2,1} -$$

$$-m_2F_{2,5} + \frac{1}{8}C_{2,1}\sqrt{10 + 2\sqrt{5}}\sqrt{5}m_4 - \frac{1}{8}C_{2,1}\sqrt{10 + 2\sqrt{5}}\sqrt{5}m_3 + \frac{1}{4}\sqrt{5}m_2F_{2,1} -$$

$$-\frac{1}{4}\sqrt{5}m_3F_{2,1} - \frac{1}{4}\sqrt{5}m_4F_{2,1} + \frac{1}{4}\sqrt{5}m_5F_{2,1} - \frac{1}{8}C_{2,1}\sqrt{10 + 2\sqrt{5}}\,m_4 + \frac{1}{8}C_{2,1}\sqrt{10 + 2\sqrt{5}}\,m_3 +$$

$$+\frac{1}{4}C_{2,1}\sqrt{10 + 2\sqrt{5}}\,m_5 - \frac{1}{4}C_{2,1}\sqrt{10 + 2\sqrt{5}}\,m_2 = 0$$

The results of solving this system using the Maple-*solve* procedure are very cumbersome. For example, for one arbitrary constant, it looks like

$$C_{2,1} = -(C_{2,3}m_5^2 + C_{2,3}m_3^2 - C_{2,3}m_4^2 + \sqrt{5}C_{2,4}m_3^2 + \sqrt{5}C_{2,4}m_4^2 + C_{2,4}m_2m_4 +$$
$$+C_{2,4}m_3m_4 + C_{2,4}m_3m_5 - 3C_{2,4}m_2m_5 + C_{2,4}m_5m_4 + \sqrt{5}C_{2,4}m_2^2 + \sqrt{5}C_{2,4}m_5^2 +$$
$$+\sqrt{5}C_{2,3}m_5m_4 + \sqrt{5}C_{2,3}m_2m_5 + \sqrt{5}C_{2,3}m_3m_5 + \sqrt{5}C_{2,3}m_3m_4 + \sqrt{5}C_{2,3}m_2m_2 +$$
$$+\sqrt{5}C_{2,3}m_2m_3 - C_{2,4}m_5^2 - C_{2,4}m_2^2 + C_{2,4}m_3^2 + C_{2,4}m_4^2 + \sqrt{5}C_{2,3}m_3^2 + \sqrt{5}C_{2,3}m_4^2 +$$
$$+\sqrt{5}C_{2,3}m_2^2 - 3\,C_{2,3}m_2m_5 + C_{2,3}m_3m_2 - C_{2,3}m_3m_5 + C_{2,3}m_4m_5 - C_{2,3}m_4m_2 +$$
$$+3C_{2,3}m_4m_3 + \sqrt{5}\,C_{2,3}m_5^2 + \sqrt{5}C_{2,4}m_2m_5 + \sqrt{5}C_{2,4}m_4m_3 + \sqrt{5}C_{2,4}m_2m_3 + \sqrt{5}C_{2,4}m_4m_5 +$$
$$+\sqrt{5}C_{2,4}m_2m_4 + \sqrt{5}C_{2,4}m_5m_3)/(2m_5^2 + 2m_3m_5 - \sqrt{5}m_2m_5 + 3m_5m_4 + \sqrt{5}m_5m_4 +$$
$$+m_2m_5 + 3m_3^2 + 3m_4^2 + \sqrt{5}m_4^2 + \sqrt{5}m_3^2 + 2m_2m_4 + \sqrt{5}m_2m_3 + \quad 2\quad m_2^2 \quad +4 \quad m_4m_3 +$$
$$2\sqrt{5}m_3m_4 + 3m_2m_3)$$

This expression requires a significant simplification that does not work automatically. The expression is a fraction. Therefore, simplification is possible when the numerator and denominator are factorized separately, and then they are automatically canceled by a common factor. These operations can be performed using a construction like this:

$$C[2,1] = \frac{factor(numer(C[2,1]))}{factor(denom(C[2,1]))}.$$

The result is a very simple expression:
$$C_{2,1} := \frac{1}{2}C_{2,5}\sqrt{5} - \frac{1}{2}C_{2,5} - C_{2,4}.$$

As follows from the theory of linear differential equations, in the resulting system of 5 equations with 10 variables, only two arbitrary variables are linearly independent. Indeed, the result of solving this system using the Maple-*solve* procedure gives the following results:

$$C_{2,1} := \frac{1}{2}C_{2,5}\sqrt{5} - \frac{1}{2}C_{2,5} - C_{2,4}$$

$$F_{2,1} := \frac{1}{2}\frac{(\sqrt{5} - 1)(3C_{2,5} + C_{2,5}\sqrt{5} + 2C_{2,4})}{\sqrt{10 + 2\sqrt{5}}}$$

$$C_{2,2} := -\frac{1}{2}(C_{2,5} + C_{2,4}) * (\sqrt{5} - 1)$$

$$F_{2,2} := -\frac{(\sqrt{5} + 1)(-C_{2,5} + C_{2,4})}{\sqrt{10 + 2\sqrt{5}}}$$

$$C_{2,3} := -\frac{1}{4}(C_{2,5} + C_{2,5}\sqrt{5} - 2C_{2,4}) * (\sqrt{5} - 1)$$

$$F_{2,3} := \frac{1}{2}\frac{(\sqrt{5} + 1)(-3C_{2,5} + C_{2,5}\sqrt{5} - 2C_{2,4})}{\sqrt{10 + 2\sqrt{5}}} \quad (13)$$

$$C_{2,4} := C_{2,4} \qquad\qquad F_{2,4} := -\frac{(\sqrt{5}-1)(C_{2,5} + C_{2,5}\sqrt{5} - C_{2,4})}{\sqrt{10 + 2\sqrt{5}}}$$

$$C_{2,5} := C_{2,5} \qquad\qquad F_{2,5} := -\frac{C_{2,5}\sqrt{5} - C_{2,5} + 2C_{2,4})}{\sqrt{10 + 2\sqrt{5}}}$$

It should be noted that the choice of independent constants is carried out by the program itself, and in different sessions their composition can change, which causes some inconvenience. But it is possible to construct a solution algorithm in such a way that these inconveniences are eliminated. This is exactly how the algorithm used in this work is constructed. Of course, it was possible to preassign these constants, but, as experience with the Maple *solve* command shows, automatic selection gives solutions of the simplest form.

Next, you need to construct 5 components of the general solution of the system of equations (3), which are the sums of particular solutions - one of the form (11) and two of the types (12), which will depend on five independent constants. To determine these constants, it is necessary to take into account the initial condition for (3): at the initial moment, all the components of the solution, except for the first one, must take a zero value, the first component must take a single value, that is, as for an exponent from a real argument. In this case, we get the following system of five equations:

$$\begin{aligned}
C_{1,5} - \tfrac{1}{2}(\sqrt{5}-1)(C_{2,4} + C_{2,3}) + \tfrac{1}{2}(1+\sqrt{5})(C_{4,4} + C_{4,3}) &= 1 \\
C_{1,5} - \tfrac{1}{4}(\sqrt{5}-1)(C_{2,4} + C_{2,4}\sqrt{5} - 2C_{2,3}) - \tfrac{1}{4}(1+\sqrt{5})(C_{4,4}\sqrt{5} - C_{4,4} + 2C_{4,3}) &= 0 \\
C_{1,5} + C_{2,3} + C_{4,3} &= 0 \qquad\qquad (14)\\
C_{1,5} + C_{2,4} + C_{4,3} &= 0 \\
C_{1,5} + \tfrac{1}{2}(\sqrt{5}C_{2,4} - \tfrac{1}{2}C_{2,4} - C_{2,3} - \tfrac{1}{2}C_{4,4} - \tfrac{1}{2}C_{4,4}\sqrt{5} - C_{4,3} &= 0
\end{aligned}$$

After solving this system, substituting the values of independent constants in (12) and (13), we obtain the values of all 25 constants, which are given in table:

|   | $C_{1i}$ | $C_{2i}$ | $F_{2i}$ | $C_{3i}$ | $F_{3i}$ |
|---|---|---|---|---|---|
| 1 | 1/5 | 2/5 | 0 | 2/5 | 0 |
| 2 | 1/5 | $\frac{1}{10}\sqrt{5} - \frac{1}{10}$ | $\frac{1}{5}\frac{5+\sqrt{5}}{\sqrt{10+2\sqrt{5}}}$ | $-\frac{1}{10}\sqrt{5} - \frac{1}{10}$ | $\frac{1}{10}\sqrt{2}\sqrt{5-\sqrt{5}}$ |
| 3 | 1/5 | $-\frac{1}{10}\sqrt{5} - \frac{1}{10}$ | $\frac{2}{5}\frac{\sqrt{5}}{\sqrt{10+2\sqrt{5}}}$ | $\frac{1}{10}\sqrt{5} - \frac{1}{10}$ | $\frac{1}{5}\frac{\sqrt{2\sqrt{5}}}{\sqrt{5-\sqrt{5}}}$ |
| 4 | 1/5 | $-\frac{1}{10}\sqrt{5} - \frac{1}{10}$ | $\frac{2}{5}\frac{\sqrt{5}}{\sqrt{10+2\sqrt{5}}}$ | $\frac{1}{10}\sqrt{5} - \frac{1}{10}$ | $\frac{1}{5}\frac{\sqrt{2\sqrt{5}}}{\sqrt{5-\sqrt{5}}}$ |
| 5 | 1/5 | $\frac{1}{10}\sqrt{5} - \frac{1}{10}$ | $-\frac{1}{5}\frac{5+\sqrt{5}}{\sqrt{10+2\sqrt{5}}}$ | $-\frac{1}{10}\sqrt{5} - \frac{1}{10}$ | $\frac{1}{10}\sqrt{2}\sqrt{5-\sqrt{5}}$ |

Now you can write the final formula for representing the exponent in the HNS $G_{51}$:

$$\text{Exp}(M) = \sum_{i=1}^{5} (\frac{1}{5} e^{\lambda_1} + e^{re\lambda_2}(C_{2i} \cos(im\lambda_2) + F_{2i} \sin(im\lambda_2)) + e^{re\lambda_4}(C_{3i} \cos(im\lambda_4) + F_{3i} \sin(im\lambda_4)) e_i, \quad (25)$$

where: $M = \sum_{i-1}^{5} m_i e_i$, $\lambda_1 = \sum_{i-1}^{5} m_i$,

$$re\lambda_{2,4} = \frac{\sqrt{10 \pm \sqrt{5}}}{32} ((m_2 + m_3 + m_4 + m_5) \pm \sqrt{5}(m_2 - m_3 - m_4 + m_5)),$$

$$im\lambda_{2,4} = \frac{\sqrt{10 \pm \sqrt{5}}}{8} ((\pm 2m_2 \mp 2m_5 \mp m_3 \pm m_4 + \sqrt{5}(m_3 - m_4)).$$

**Conclusions**

1. The method considered in the work allows one to construct representations of the exponential in the HNS of the fifth dimension. This makes it possible to synthesize isomorphic HNS with a lower fullness of the Keli tables, which, in turn, makes it possible to increase the efficiency of computational processes in the original HNS.

2. The structure of the method is such that the method can be generalized to higher-dimensional HNS, which is important for practical applications, for example, in the synthesis of high-dimensional digital filter structures.

**REFERENCES**

1. HamiltonW. R. Researches respecting quaternions: first series//Transactions of the Royal Irish Academy, 1848,21, part 1, P. 199–296.
2. Brackx F. The exponential function of a quaternion variable// Applicable Analysis, 1979, 8, P. 265– 276.
3. Scheicher K., Tichy R. F., Tomantschger K. W.,Elementary Inequalities in Hypercomplex Numbers Anzeiger, 1997, Abt. II, No 134,p. 3–10.
4. КалиновскийЯ.А., РоенкоН.В.,. Синьков М.В Методы построения нелинейностей в расширениях комплексных чисел//Кибернетика и системный анализ, 1996, № 4,С. 178–181.
5. СиньковМ.В., КалиновскийЯ.А., РоенкоН.В. Building nonlinear functions in quaternion and other hypercomplex number systems for the solution of applied mecanics problem // Proc. of the First Int. Conf. On parallel processing and appl. Math.,Poland, 1994,P. 170–177.
6. СиньковМ.В., КалиновскийЯ.А., БояриноваЮ.Е., ФедоренкоА.Ф.О дифференциальных уравнениях, определяющих функции гиперкомплексного переменного //Реєстрація, зберігання і обробкаданих, том 8 № 3, 2006, стр.20 -24.
7. Каліновський Я.О., Синьков М. В., Боярінова Ю.Є., Федоренко О.В. Зображеннянелінійностей в скінченновимірнихгіперкомплекснихчислових системах //Доповіді НАНУ, 2008.№8,С.43-51.
8. КалиновскийЯ.А., Синьков М. В., Бояринова Ю.Е. Конечномерные гиперкомплексные числовые системы. Основы теории. Применения// К.: Инфодрук, 2010, 388с.
9. Клименко В.П., Ляхов А.Л., Гвоздик Д.Н.Современные особенности развития систем компьютерной алгебры // Математичнімашини і системи,2011,№ 2, С. 3 - 18.
10. О.Татарников. Обзорпрограмм для символьной математики .[Электронный ресурс]. Режим доступа: https://compress.ru/article.aspx?id=16152
11. Калиновский Я.А.,.БояриноваЮ.Е, ХицкоЯ.В.Гиперкомплексные вычисления в Maple // ИПРИ НАНУ, 2020. 180с. ISBN 978-966-02-8879-9.
12. Кострикин А.И. Введение в алгебру/М.:Наука, 1977.-496с.
13. Калиновский Я.А.,.БояриноваЮ.ЕВысокоразмерные изоморфные гиперкомплексные числовые системы и их применения// К.: ИПРИ НАНУ, 2012, 183с.
14. Корн Г. Справочник по математике для научных работников и инженеров// М.: Наука, 1974, 832 с.

REFERENCES

1. HamiltonW. R. Researches respecting quaternions: first series // Transactions of the Royal Irish Academy, 1848, 21, part 1, P. 199–296.
2. Brackx F. The exponential function of a quaternion variable // Applicable Analysis, 1979, 8, P. 265– 276.
3. Scheicher K., Tichy R. F., Tomantschger K. W. // Elementary Inequalities in Hypercomplex Numbers Anzeiger, 1997, Abt. II, No 134, S. 3–10.
4. KalinovskiyYa.A.,RoenkoN.V., Sinkov M.V. Methods for constructing nonlinearities in extensions of complex numbers // Cybernetics and Systems Analysis, 1996, № 4 ,P. 178–181.
5. KalinovskiyYa.A.,RoenkoN.V., Sinkov M.V. Building nonlinear functions in quaternion and other hypercomplex number systems for the solution of applied mecanics problem// Proc. of the First Int. Conf. On parallel processing and appl. Math.,Poland, 1994, P. 170–177.

6. SinkovM.V., KalinovskiyYa.A.,BoiarinovaYu.E., FedorenkoA.F.On differential equations defining functions of a hypercomplex variable //Data Recording, Storage & Processing, tom 8 № 3, 2006, p.20 -24.
7. SinkovM.V., KalinovskiyYa.A.,BoiarinovaYu.E., FedorenkoA.F.Imaging non-linearities in scanned-dynamic hypercomplex number systems //Dopovidi NANU,- 2008.-№8.-P.43-51.
8. SinkovM.V., BoiarinovaYu.E., KalinovskiyYa.A.Finite-dimensional hypercomplex number systems. Foundations of the theory. Applications// K.: Infodruk, 2010, 388p.
9. Klimenko V.P., Lyahov A.L., Gvozdik D.NModern features of the development of computer algebra systems // Matematicalmachinesandsistems, 2011, № 2, p. 3 – 18
10. O.Tatarnikov. Overview of Symbolic Mathematics Programs.[Electronic resource]. Access mode: https://compress.ru/article.aspx?id=16152
11. KalinovskiyYa.A, BoiarinovaYu.E., HitskoYa.V. Hypercomplex calculates in Maple// IPRI NANU, 2020. 180p. ISBN 978-966-02-8879-9.
12. Kostrikin A.I. Introduction to algebra /M.:Nauka, 1977,496p.
13. KalinovskiyYa.A, BoiarinovaYu.EHigh-dimensional isomorphic hypercomplex number systems and their applications //K.: IPRI NANU, 2012,183p.
14. Korn G. A guide to mathematics for scientists and engineers// M.: Nauka, 1974, 832 p.

# Method for representing an exponent in a fifth-dimensional hypercomplex number systems using a hypercomplex computing software

The paper presents a method for determining the isomorphism of hypercomplex number systems (HNS), which is based on the analysis of representations of exponential functions in the normalized form of isomorphic HNS. The method allows to obtain an explicit form of the isomorphism operator, which can be used to model highorder digital reversible filters. The calculations given in the text are made by means of the hypercomplex computing software (HCS) developed by authors that considerably simplifies them.

Digital filters can be implemented both by soft ware and with the help of specialized equipment. The use of hypercomplex number systems for the construction of digital filter structures by the software method can provide significant advantages. Digital filters with hypercomplex parameters have higher speed and better characteristics in terms of total parametric sensitivity.

When synthesizing the structures of digital filters with hypercomplex parameters, it is necessary to use HNS with completely filled multiplication tables. This is due to the fact that such HNS provide the synthesis of any transfer function of the filter. But algebraic operations with such hypercomplex number srequire a large number of operations on real numbers. Thus, when using canonical HNS dimensions $N$, the multiplication of two hypercomplex numbers is reduced to $N^2$ multiplication and $N(N-1)$ addition of real numbers. And if noncanonical HNS are used, the number of operations is even greater (to $N^3$ and $N(N-1)^2$ respectively).

At the same time, it is known that an effective method of reducing the volume of calculations is the transition to isomorphic weakly filled HNS type direct products of HNS of smaller dimensions. For example, the multiplication of two numbers in a quadriplex HNS K requires 16 real operations, and in an isomorphic bicomplex HNS $C \oplus C$ only 6.

But the construction of isomorphic HNSs, especially high dimensions, causes certain algorithmic and computational difficulties.

To synthesize the structures of digital filters, it is necessary to have three objects: two isomorphic HNS $\Gamma_1 \simeq \Gamma_2$ and an isomorphism operator $L$. All three objects are interconnected. When synthesizing the structure of a particular filter, it is known in advance that one of the HNS must be a direct product of the HNS R and C, the other - a densely filled canonicalor non-canonical HNS. Moreover, it is better to use the canonical HNS, because it greatly simplifies the process of synthesis of the filter structure and the study of its sensitivity. The dimension of b oth HNS should be equal to the order of the filter.

Thus, the purpose oft his work is to create and study a method for generating pairs of isomorphic HNS of a given dimension, which includes the construction of the isomorphism operator of these HNS.

*K e y w o r d s: hypercomplex number system, representation of functions, exponent, characteristic number, computer algebra systems, algebraic operation, Kely table.*

КАЛІНОВСЬКИЙ Яків Олександрович, д-р техн. наук, ст. наук. співроб. Інституту проблем реєстрації інформації НАН України. У 1965 р. закінчив Київський політехнічний інститут. Область наукових досліджень – математичне моделювання, теорія та застосування гіперкомплексних числових систем.

KALINOVSKY Yakiv Oleksandrovych, Doctor of Technical Sciences, Senior Researcher. Institute of Information Registration Problems of the National Academy of Sciences of Ukraine. In 1965 he graduated from Kyiv Polytechnic Institute. Area of research - mathematical modeling, theory and application of hypercomplex number systems.

БОЯРІНОВА Юлія Євгенівна, канд. техн. наук, доцент кафедри Системного програмування та спеціалізованих комп'ютерних систем НТУУ «Київський політехнічний інститут ім. Ігоря Сікорського». У 1997 р. закінчила НТУУ «Київський політехнічний інститут ім. Ігоря Сікорського». Область наукових досліджень – теорія та застосування гіперкомплексних числових систем при математичному моделюванні.

BOIARINOVA Yuliia Yevhenivna, PhD, Associate Professor of Systems Programming and Specialized Computer Systems National Technical University of Ukraine "Igor Sikorsky Kyiv Politechnic Institute".In 1997 she graduated from National Technical University of Ukraine "Igor Sikorsky Kyiv Politechnic Institute".Area of research - theory and application of hypercomplex number systems in mathematical modeling